\documentclass[fleqn,12pt,twoside]{article}
\usepackage{espcrc1}
\usepackage[latin1]{inputenc}

\usepackage{graphicx}
\usepackage[figuresright]{rotating}

\newcommand{\Zmoy}{\langle Z \rangle}

\newcommand{\AmS}{{\protect\the\textfont2
  A\kern-.1667em\lower.5ex\hbox{M}\kern-.125emS}}
\DeclareMathAlphabet{\bi}{OML}{cmm}{b}{it}

\title{Liquid-gas phase transition in hot nuclei studied with INDRA}

\author{B.~Borderie\address[IPNO]{Institut de Physique Nucl\'eaire, 
                 IN2P3-CNRS, F-91406 Orsay Cedex, France},
	R.~Bougault\address[LPC]{LPC Caen, IN2P3-CNRS/ENSICAEN et
                 Universit\'e, F-14050 Caen cedex, France},
        P.~D\'esesquelles{\addressmark[IPNO]},
        E.~Galichet{\addressmark[IPNO]},		 
        B.~Guiot\address[GANIL]{GANIL, DSM-CEA/IN2P3-CNRS, B.P. 5027,
	         F-14021 Caen Cedex, France}
		 \thanks{Present address: INFN, Sezione di
                 Bologna and Dipartimento di Fisica, Universit\`a di Bologna,
		 Italy},
        Ph.~Lautesse\address[IPNL]{Institut de Physique Nucl\'eaire, 
                 IN2P3-CNRS
                 et Universit\'e, Villeurbanne, France},
	N.~Le~Neindre{\addressmark[IPNO]},
	J.~Marie{\addressmark[LPC]},
	M.~P\^arlog\address[NIPNE]{ National Institute for Physics and Nuclear
	          Engineering, Bucharest-M\u{a}gurele, Romania},
	M.~Pichon{\addressmark[LPC]},
	M. F. Rivet{\addressmark[IPNO]},
	E.~Rosato\address[NAP]{Dipartimento di Scienze Fisiche e Sezione INFN,
                  Universit\'a di Napoli, Napoly,
                  Italy},
        G.~T\u{a}b\u{a}caru\addressmark[NIPNE]
                  \thanks{Present address: Cyclotron Institute,
                  Texas A\&M University, College Station, Texas 77845, USA},
        M.~Vigilante{\addressmark[NAP]}
        and
	J.~P.~Wieleczko{\addressmark[GANIL]}}

\begin{document}

\maketitle

\begin{abstract}
Thanks to the high detection quality of the INDRA array, signatures
related to the dynamics (spinodal decomposition) and thermodynamics
(negative microcanonical heat capacity) of a liquid-gas phase transition
have been simultaneously studied in multifragmentation events in the Fermi
energy domain.
The correlation between both types of signals 
strongly supports
the existence of a first order phase transition for hot nuclei.
\end{abstract}

\section{Introduction}
The main goal of multifragmentation studies concerns its relation 
with subcritical and/or critical phenomena. Thus it is fully connected to
the nature of the phase transition of nuclear matter which is expected of
the liquid-gas type due to the specific form of the nucleon-nucleon
interaction. Despite the large number of experimental results, fundamental
questions could not be addressed for a long time. We are facing an exciting
but also very complex subject where only very sophisticated detector arrays
like INDRA~\cite{I3-Pou95,I5-Pou96} can produce the high quality experiments
needed~\cite{MDA00,Hau00,Elli00,Xu00,I46-Bor02,MDA02,Radu02,Kle02,Ell02}.
From collisions between
nuclei we hope to reveal a phase transition for those finite
objects and derive the related equation of state of nuclear matter.
In parallel to the experimental effort a big theoretical effort started
a few years ago to characterize and propose signatures of phase transitions
in finite systems~\cite{Gro01,Cho02,Gul03}. Phase transitions should be
reconsidered  from a
more general point of view, out of the thermodynamical limit. Thus,
experiments on nuclear multifragmentation are also used as test-bench
for the elaboration of the statistical physics of finite systems.

The paper describes some major recent results obtained with the INDRA array. 
It is organized as follows. After some information concerning the
description of collisions and associated simulations, the dynamics of phase
transition in
hot nuclei and the related proposed signature will be presented.
Then specific phase transition signals related to the thermodynamics of 
finite systems will be discussed in terms of robustness. 
Finally, from systematic studies, the coherence between signals will be
demonstrated.  

\section{Description of collisions and multifragmentation}
Semi-classical simulations of nucleus-nucleus collisions based on   
the nuclear Botzmann
equation (VUU,LV,BUU,BNV)~\cite{Kru85,Gre87,Ber88,Bon94},                   
were very successful in reproducing a variety of experimental dynamical    
observables. However
as they follow the time evolution of the one body       
density, considering only the average effect of collisions between particles
(two-body collisions), they ignore               
fluctuations about the main trajectory of the                                  
system (deterministic description), which becomes a severe                     
drawback if one wants to describe processes involving instabilities,           
bifurcations or chaos. Such approaches are appropriate
during      
the early stages of nuclear collisions, when the system is hot and
compressed, but they become inadequate when expansion and cooling have brought
the system in the instability (spinodal) zone of the phase diagram. In
such a
scenario, it is essential to include the fluctuations. This is done in
Quantum Molecular        
Dynamics methods and in the Stochastic Mean Field (SMF) approach.  
We shall now focus    
on SMF approach, a natural evolution of deterministic semi-classical
simulations, which was compared to INDRA data. 

\subsection{Brownian one-body dynamics (BOB)}
In BOB simulations, fluctuations in one-body dynamics are introduced by
employing a Brownian force in the kinetic equations.
 The starting       
point is the so-called Boltzmann-Langevin equation (BLE):                     
\begin{equation}                                                               
\frac{\partial f}{\partial t} = \{h[f],f\} +  \overline{I}[f] + \delta I[f]    
\end{equation}                                                                 

which was introduced for heavy-ion collisions in references~\cite{Ayi88,Ran90}.
$f$ is the one-body phase space density. The first term on the r.h.s.
produces the collisionless
propagation of $f$ in the self-consistent one-body field described by the
effective Hamiltonian. The second term (second source of evolution called      
collision term)                                                                
represents the average effect of the                                           
residual Pauli-suppressed two-body collisions; this is the term included in    
LV,BUU and BNV simulations. The third term is the Langevin term which          
accounts for the fluctuating part of the two-body collisions.                  
Exact numerical solutions of the BLE are very difficult to obtain and          
have only be calculated for schematic cases in one or two                      
dimensions~\cite{Cho91}. Therefore various approximate treatments of the BLE   
have been developed.
The basic idea of BOB~\cite{Cho94} is to replace the       
fluctuating term  by                                                           
\begin{equation}                                                               
\delta\tilde{I} [f] = - \delta \bi{F} [f] . \frac{\partial f}{\partial         
\bi{p}}                                                                        
\end{equation}                                                                 
where $\delta \bi{F} (\bi{r},t)$  is the associated Brownian force             
($<\delta \bi{F}>=0$). Since the                                               
resulting Brownian one-body dynamics mimics the BL evolution, the stochastic   
force is assumed to be local in space and time. The strength of the force      
is adjusted to reproduce the growth of the most unstable modes for infinite    
nuclear matter in the unstable spinodal region (see next section).

\subsection{Comparison with data}
The comparison was made for two very heavy fused systems produced in Xe+Sn
and  
Gd+U  central collisions which undergo multifragmentation with the same
available excitation energy ( $\sim$ 7 MeV per nucleon).      
Stochastic mean-field simulations were performed for head-on collisions,
thus neglecting shape effects.  
Ingredients of simulations are detailed in~\cite{I29-Fra01}. 
After BOB simulation a second step, starting from the phase space configuration
of the primary fragments as given by BOB, followed the fragment statistical
deexcitation while preserving          
space-time correlations. Finally the events were filtered                      
to account for the experimental device.
These complete simulations very
well reproduce the observed multiplicity and charge distributions
of fragments~\cite{I29-Fra01}.                                   
Particularly the independence of the charge distribution against the mass of   
the system experimentally observed is recovered~\cite{I12-Riv98}. 
More detailed comparisons also
show a good agreement~\cite{I29-Fra01,TabBol00}.
Kinetic properties of fragments are rather well accounted for
especially for the Gd+U system. For Xe+Sn, the calculated energies fall $\sim$
20\% below the measured values which remains satisfactory if one remembers     
that there were no adjustable parameters in the simulation.

\subsection{What have we learnt ?}
From the good agreement between dynamical       
simulations and data a complete scenario is proposed~\cite{I29-Fra01}.
After a gentle compression ($\sim$1.25$\rho_{0}$), maximum at 40 fm/c,
systems expand    
and enter the unstable (negative compressibility) spinodal region at around
80 fm/c. Slightly later ($\sim$ 100     
fm/c) thermal equilibrium is achieved ($<T>$=4 MeV) at low                     
density ($\sim$0.4$\rho_{0}$) and systems                                      
have lost $\sim$5\% of their initial masses by emitting
preequilibrium particles. The radial velocities at the
surface ($\sim$0.1c) reveal the expansion of the systems and density
fluctuations (spinodal instabilities) have time to develop leading to the
formation of fragments. Around
 250 fm/c after the beginning of collisions the fragments do not
 interact any more, they are in thermal equilibrium and still bear an average
 excitation energy of $\sim$3 MeV per nucleon:   
 it is the freeze-out time with the associated freeze-out volume, it
 corresponds to the end of these BOB simulations.

\section{Dynamics of the phase transition and related signal}

Thermodynamics describes phase transitions in terms of ``static conditions''.
Information on existence and coexistence of phases are
derived depending on thermodynamical parameters
(temperature, pressure,\ldots ). How to pass from a phase to another ? What
is the time needed ? To answer these questions, dynamics of phase transitions
must be studied. A phase transition is a dynamical phenomenon with its
proper kinetics. The aim of this section is to discuss the dynamics of the 
phase transition which can be involved in hot unstable nuclei.
Two mechanisms are candidates.
(i) Nucleation which is associated with a transport of matter; diffusion is the
key mechanism which controls the kinetics; we are in presence of heterogeneous
fluctuations. (ii) Spinodal decomposition
which occurs when a system spontaneously develops
local fluctuations of concentration or density which grow exponentially
with time; characteristic time is related to the most
unstable modes. In contrast with nucleation, we are here dealing with
homogeneous fluctuations.
Given the agreement discussed  in the previous section, we will now focus on
spinodal decomposition.

\subsection{Spinodal decomposition: from nuclear matter to nuclei}
In the spinodal region some modes are amplified
because of the instability. Their associated characteristic times are almost
identical, around 30- 50 fm/c,   
depending on density ($\rho_{0}/2$-$\rho_{0}/8)$ and
temperature (0-9MeV)~\cite{Colo97,Idi94}.  
Within this scenario a breakup into nearly
equal-sized ``primitive'' fragments should be favoured in relation to the
wawelengths of the most unstable modes~\cite{Ayi95}. However this simple
picture is predicted to be strongly blurred by several effects: beating
of different modes, coalescence of nascent fragments, secondary decay of
excited fragments and, above all, finite size effects~\cite{Jac96}.
Therefore only a weak proportion of multifragmentation events with nearly
equal-sized fragments is expected.

\subsection{The related fossil signal: enhanced production of nearly
equal-sized fragments}
A few years ago a new method called higher order charge
correlations~\cite{Mor96}, which has been recently improved~\cite{Des02},
was proposed to enlighten any extra production of     
events with specific fragment partitions. The high sensitivity of the method
makes it particularly appropriate to look for small numbers of events as those
expected to have kept a memory of spinodal decomposition properties.
Thus, such a charge correlation method allows to examine model independent
signatures that would indicate a preferred decay into a number of equal-sized
fragments in events from experimental data or from simulations.
 All fragments ($Z \geq 5$) of one event with fragment    
multiplicity $M = \sum_Z n_Z$, where $n_Z$ is the number of fragments
with charge $Z$ in the partition,  are taken into account. By means of the
normalized first and second order moments of the fragment charge 
distribution in the event:
\begin{equation}
        \Zmoy = \frac{1}{M} \sum_Z n_Z Z \; , \; \; \;
        \sigma_Z^2 = \frac{1}{M} \sum_Z n_Z (Z - \Zmoy)^2
\label{equ1}
\end{equation}
%
%
one may define the higher order charge correlation function:
\begin{equation}
\left. 1+R(\sigma_Z, \Zmoy)=\frac{Y(\sigma_Z, \Zmoy)}{Y'(\sigma_Z, \Zmoy)}
\right| _{M}
\label{equ3}
\end{equation}
Here, the numerator $Y(\sigma_Z, \Zmoy)$ is the yield of events with given
$\Zmoy$  and $\sigma_Z$ values and the evaluation of the denominator
(uncorrelated yield) is deduced from the exact multinomial formula under
total charge conservation constraint (see ref.~\cite{Des02,I40-Tab02} for
details).

\begin{figure}[htb]
\begin{minipage}[c]{.45\textwidth}
\centering
\includegraphics[width=\textwidth]
{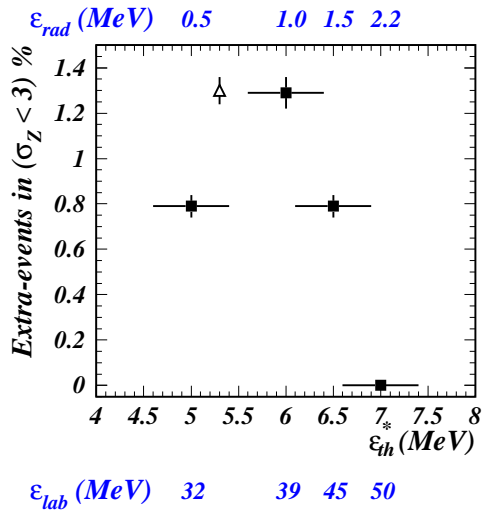}
\end{minipage}
\hspace{.05\textwidth}%
\begin{minipage}[c]{.45\textwidth}
\centering
\caption{Abnormal production of events with nearly equal-sized fragments
( $\sigma_{Z} < 3$ corresponds to the upper limit of the islet of
peaks - see~\cite{I40-Tab02} for details)
as a function of thermal
excitation energy (deduced from SMM): full points. The incident and radial
energy scales are also indicated.
The open point refers to the result from BOB simulations; the average
thermal energy is used.}%
\label{fig8}
\end{minipage}
\end{figure}

We shall now discuss the results obtained for fused systems produced in
 $^{129}$Xe central collisions on $^{nat}$Sn at four incident energies (32,
39, 45 and 50 AMeV)~\cite{I40-Tab02}. At 32 AMeV the present analysis fully
confirms the previous one obtained with the original method~\cite{I31-Bor01}
and the extra-percentage of events with nearly equal-sized fragments is
maximum at 39 AMeV incident energy. The excitation function is displayed in
fig.~\ref{fig8}.
Information on the associated thermal excitation energies (and extra radial
collective energy) involved over the incident energy domain studied can be
provided by the SMM model~\cite{Bon95} which well describes static and dynamic
observables of fragments. Starting from a freeze-out volume fixed at three
times the normal volume, the thermal excitation energies of the dilute
and homogeneous system, extracted from SMM, vary from 5.0 to 7.0 AMeV and
the added radial expansion energy remains low: from 0.5 to 2.2
AMeV~\cite{T16Sal97,T25NLN99}.
A rise and fall of the percentage of ``fossil partitions'' from spinodal
decomposition is measured. Results from BOB simulations ( head-on
$^{129}$Xe on $^{119}$Sn collisions at 32 AMeV) are also shown;
although all events in the
simulation arise from spinodal decomposition, only a very small fraction
of the final partitions ($\sim$ 1\%)  have kept nearly equal-sized fragments.
Mainly due to finite size effects the signature of spinodal    
decomposition can only reveal itself as a weak ``fossil'' signal.   

As a conclusion of that section we can say that, supported by theoretical
simulations, we              
interpret our data as a signature of spinodal instabilities as
the    
origin of multifragmentation of those fused systems in the Fermi energy domain.
Spinodal decomposition describes the dynamics of a first order phase
transition, the present observations thus support the existence of such a
transition for hot finite nuclear matter.

\section{Thermodynamics of finite systems and related signals}
In section 2 we have shown that experimental   
charge and multiplicity distributions were reproduced with BOB simulations
while average fragment kinetic energies were accounted for within 20 \%.
These same
data were also well accounted for with the statistical model
SMM~\cite{T16Sal97,T25NLN99,LenBol00}.  From these two findings
 we can deduce that the dynamics involved is
sufficiently         
chaotic to finally explore enough of the phase space in order to describe
fragment      
production through a statistical approach. Selected events correspond to
``statistical samples'' (which have nothing to do with ergodic arguments)
and therefore we can go further and use statistical mechanics of finite
systems.

\subsection{Different signals and their robustness}
From statistical physics of finite systems three direct signals of a first
order phase transition are proposed: (i) the caloric curve at constant
pressure,
which is predicted to present a negative curvature, (ii) the
presence
of negative microcanonical heat capacity~\cite{Cho02} and (iii)
the existence of a bimodality of the event distribution as a function of an
 order parameter~\cite{Cho01}.
They are all related to a curvature
anomaly of a thermodynamical potential due to surface effects in the
unstable region of coexistence.

Do those signals have equivalent robustness ?
 Experimentally one does not explore the caloric curves
at constant pressure nor at constant volume. In fact many different measured
caloric
curves can be generated from experiments depending on the path followed on 
 the microcanonical equation of state
surface (T versus excitation energy and average volume).
Therefore
signals from shapes of caloric curves ( plateau-like) appear not as robust
as 
the two others signals. The
method proposed to measure the
microcanonical heat capacity uses partial energy fluctations. Abnormal
large fluctuations (as compared to the canonical expectation) are predicted
 to be always seen, independently of the path, if microcanonical negative
heat capacity is present~\cite{Cho00}. Finally bimodality is predicted to
be very robust.

\subsection{Abnormal kinetic energy fluctuations and negative
microcanonical heat capacity}
\begin{figure}[!hbt]
\begin{minipage}[c]{0.45\textwidth}
\centering
\includegraphics[width=\textwidth]
{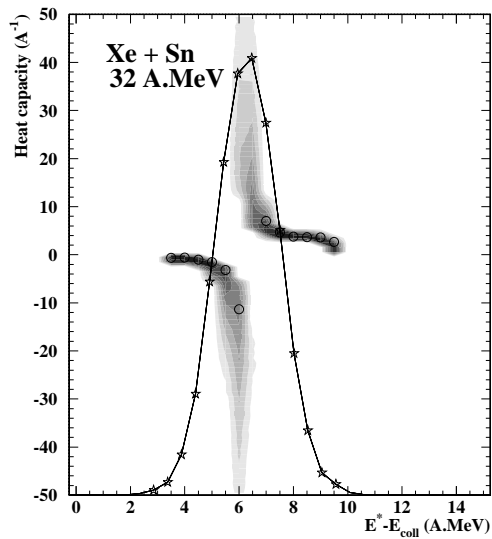}
\end{minipage}%
\hspace{.05\textwidth}%
\begin{minipage}[c]{0.45\textwidth}
\centering
\caption{
Microcanonical heat capacity per nucleon as a function of the excitation
energy (corrected in average for the radial
collective energy) measured in central collisions between Xe and Sn.
From~\cite{NLNBorm00}}%
\label{fig:b5}
\end{minipage}
\end{figure}

The prescription proposed to estimate microcanonical heat capacity is
based on the fact that
 for a given total energy of a system, the
average partial energy stored in a part of the system is a good microcanonical
thermometer, while the associated fluctuations can be
used to construct the heat capacity~\cite{Cho02}.
From experiments the most simple decomposition of the total energy $E^{*}$
is in
a kinetic part $E_{1}$ and a potential part $E_{2}$ (Coulomb energy + total
mass excess).
However these quantities have to be determined at freeze-out  and
consequently it is necessary to trace back this
configuration on an event by event basis.
This needs, in principle, the knowledge of the freeze-out
volume and of all the particles evaporated from primary hot
fragments including the undetected neutrons. Consequently
some reasonable working hypotheses are used, eventually constrained by
specific experimental results~\cite{MDA02}.
An example of results obtained from INDRA data is displayed in 
fig.~\ref{fig:b5}.
\section{Observation of correlated signals}
The spinodal region is also thermally unstable.         
This property has a well known consequence at the thermodynamical
limit: the condition of maximal entropy (Maxwell construction)
fixes a constant temperature for a given isobar
in the coexistence region. But for finite systems the entropy
presents a convex dip in that region and
a direct consequence is the occurrence of a negative heat capacity.
Thus, when spinodal decomposition is observed, one
must measure a negative heat capacity.
The reverse is not true because spinodal decomposition needs
sufficient time to occur (about 100 fm/c, see subsection 3.1).
Thus, systems may penetrate the spinodal region and multifragment through
nucleation rather than spinodal decomposition.

Therefore both signals (spinodal decomposition and negative microcanonical
heat capacity) have been simultaneously studied on different fused systems
which undergo multifragmentation~\cite{T25NLN99,I40-Tab02,T32Gui02}.
Results are summarized in table\ref{table1}.
\begin{table*}[htb]
\caption{Summary of the findings for phase transition signals.}
\begin{tabular}{c|ccccccc}
system & Ni+Au & Ni+Ni & Xe+Sn & Xe+Sn & Ni+Au & Xe+Sn & Xe+Sn \\
\hline
Incident Energy\\AMeV & 32. & 32. & 32. & 39. & 52. & 45. & 50. \\
\hline
Thermal energy\\AMeV& 5.0$\pm$1.0 & 5.0$\pm$1.0 & 5.0$\pm$0.5 & 6.0$\pm$0.5
& 6.0$\pm$1.0 & 6.5$\pm$0.5 & 7.0$\pm$0.5 \\                
\hline
Radial energy\\AMeV& 0. & 0.8$\pm$0.5 & 0.5$\pm$0.2 & 1.0$\pm$0.3 & 0. &
1.5$\pm$0.4 & 2.2$\pm$0.4 \\                   
\hline   
Spinodal\\decomposition & no & yes? & yes & yes & yes & yes & no \\         
\hline
Negative microc.\\heat capacity & no & yes& yes & yes & yes & yes? & no \\
\end{tabular}
\label{table1}
\end{table*}
For the different systems we have also indicated the associated thermal and
radial collective energies extracted (see subsection 3.2. and
reference~\cite{T32Gui02} for details). We observe a correlation between
the two signals, which strongly supports the presence of a first order phase
transition for hot nuclei undergoing multifragmentation in the Fermi energy
domain. Moreover this correlation reinforces the fact that spinodal
decomposition (even if evidenced by a small fossil signal) is indeed the
dynamics of the phase transition.
Both signals are present when a total (thermal+radial) energy in the range
5.5-7.0 AMeV is measured. Note that the effect of a very gentle compression
phase leading to 0.5-1.0 AMeV radial expansion energy plays the same role
as a slightly higher thermal energy (Ni+Au system at 52 AMeV). This can be
understood in terms of a required threshold for expansion energy; in the
latter case this threshold should be reached by thermal expansion only.

\section{Other signals}

Several other signals are under studies using INDRA data.
The so called Fisher scaling is observed on a large set of central (fused
systems) and peripheral collisions~\cite{NLNBorm02}. It was tentatively used in
reference~\cite{Ell02}
to derive
information on the critical point of finite neutral nuclear matter;
we notice that, for the fused
systems previously discussed, the observed ``pseudo'' critical point
appears inside the coexistence zone (presence of the two correlated signals
discussed in the previous section).
This last remark is also valid for analysis of data within the universal
fluctuations framework
(see~\cite{Bot01} - fluctuations of the size of the heaviest fragment).
Within this theory, transition from an ordered phase to a disordered phase
can be well identified through the largest fluctuations nature provides for
an order parameter.
Finally one can make the more general comment that for finite
nuclei the largest
fluctuations measured (kinetic energy and size of the heaviest fragment) are
associated to the coexistence region (see also~\cite{Gul03}).

Concerning bimodality, encouraging
results have been recently obtained using data from peripheral Au+Au collisions
studied with INDRA at GSI~\cite{PicBorm03}. A correlation between
bimodality and negative
microcanonical heat capacity is observed.

\section{Conclusions and perspectives}
The correlation observed between the presented signals completely argues
for the existence
of a liquid-gas type phase transition in hot nuclei. The nature of the
dynamics of the phase transition has been evidenced; spinodal instabilities
are thus shown for the first time in a finite system. Many systematic studies
including correlations between different signals, independent measurements to
characterize systems at freeze-out, are now required for establishing a
metrology
of the phase diagram and the related equation of state for hot nuclei and 
nuclear matter. The introduction of the N/Z degree of freedom appears also
for the future as a second very exciting challenge.


\end{document}